**Evaluating Cognitive and Neuropsychological Assessments - A Comprehensive Review**

Chuang Li[1,2], Rubing Lin[3], Yantong Liu[4], Yichen Wei[5,*]

1 Department of Biological Sciences, Purdue University, West Lafayette, IN, 47906, USA

2 College of Veterinary Medicine, China Agricultural University, Beijing, 100193, China

3 Department of Orthopedics, Shenzhen Children's Hospital, Guangdong, 518000, China

4 Department of Computer and Information Engineering, Kunsan National University, Gunsan, 54150, Republic of Korea

5 School of Life and Environmental Sciences, The University of Sydney, Sydney, NSW 2006

* Correspondence: Yichen Wei

## Abstract

Cognitive impairments in older adults represent a significant public health concern, necessitating accurate diagnostic and monitoring strategies. In this study, the principal cognitive and neuropsychological evaluations employed for the diagnosis and longitudinal observation of cognitive deficits in the elderly are investigated. An analytical review of instruments including the Mini-Mental State Examination (MMSE), Digit Symbol Substitution Test (DSST), Montreal Cognitive Assessment (MoCA), and Trail Making Test (TMT) is conducted. This examination encompasses an assessment of each instrument's methodology, efficacy, advantages, and limitations. The objective is to enhance comprehension of these assessments for the early identification and effective management of conditions such as dementia and mild cognitive impairment, thereby contributing to the advancement of cognitive health within the geriatric population.

# Introduction

The aging process is accompanied by various degrees of cognitive decline, impacting the quality of life and independence of older adults [1, 2]. As the life-expectancy increases, it is important to properly measure the cognitive ability in elderly. Cognitive and neuropsychological assessments provide important tools for assessing, diagnosing, and monitoring cognitive impairments, including dementia [3], Alzheimer's disease [3, 4], and mild cognitive impairment (MCI) [5, 6]. These assessments range from broad screenings that provide an overview of cognitive health to specific tests that evaluate cognitive domains such as memory, attention, executive function, and visuospatial skills.

Among the extensive tools available, the Mini-Mental State Examination (MMSE) [7], Digit Symbol Substitution Test (DSST) [8, 9], Montreal Cognitive Assessment (MoCA) [10], and Trail Making Test [11] are widespread use in clinical and research. Each of these assessments has its unique methodology, strengths, and limitations, making them suitable for different contexts and objectives.

This paper aims to provide a description and overview of these common cognitive and neuropsychological assessments used in older adults. By introducing their methodology, significance, and the context in which they are most effectively employed, the paper illustrates the roles of these tests in the early detection and management of cognitive decline in the aging population. Through a comparative analysis, we will explore the advantages and disadvantages of these tests, ensuring their efficient use to inform cognitive health in older adults.

# 1. Mini-Mental State Examination (MMSE)

The MMSE is a brief 30-point questionnaire that is used extensively in clinical and research settings to measure cognitive impairment. It was introduced by Folstein et al. in 1975 and has since become a standard tool for cognitive evaluation [7]. The MMSE tests various cognitive functions including arithmetic, memory, and orientation. Its primary objectives are to screen for cognitive impairment, to estimate the severity of cognitive dysfunction, and to monitor cognitive changes over time, particularly in the elderly. Administering the MMSE involves a series of questions and simple tasks encompassing different cognitive domains. The administration time is typically around 5-10 minutes. The test includes tasks such as asking the patient to recall a list of words, perform basic calculations, and identify the current date and location . The maximum score is 30 points, with a score of 23 or lower typically indicating cognitive impairment. However, the cutoff points can vary based on education and demographic factors.

## 1.1 Strength and limitations

One of the main advantages of the MMSE is its brevity and ease of use, which makes it a convenient tool for both clinicians and researchers. It requires no special equipment and can be administered in a variety of settings. Furthermore, its wide recognition and standardization allow for comparison across different populations and studies [12].

The simplicity of the MMSE can also be a limitation [13]. It may not detect less obvious forms of cognitive impairment, particularly in the early stages, or in individuals with higher education or intelligence, who may score within normal ranges despite cognitive decline [14, 15]. The limitations of MMSE also include poor test-retest reliability, limited sensitivity to subtle brain abnormalities, and poor construct validity, which restrict its utility in diagnosing and monitoring cognitive impairments accurately [16].

## 1.2 Applications of MMSE

MMSE scores contribute significantly to clinical research aimed at understanding the natural history and trajectory of cognitive decline [17-19]. Through the MMSE, researchers can identify patterns and rates of cognitive changes, supporting the development of clinical models for predicting the progression of cognitive impairments. This facilitates early diagnosis and the opportunity to intervene before significant deterioration occurs.

In studies investigating the outcomes of various interventions—ranging from pharmacological to non-pharmacological approaches—the MMSE is used to quantify cognitive benefits. For example, a study found that weight loss combined with certain plasma biomarkers was associated with higher cognitive decline over a four-year period, as assessed by the MMSE [20]. In clinical trial setting, a study, investigated the impact of Acetyl-L-carnitine (ALCAR) treatment in pre-frail older patients on cognitive function, where MMSE scores were used to document improvements in memory and cognitive processes [21-23].

# 2. Digit Symbol Substitution Test (DSST)

The DSST requires participants to match symbols to corresponding digits according to a key within a specified time limit, usually 90 to 120 seconds. The task involves a simple substitution cipher where each digit (from 1 to 9) is paired with a unique symbol. Participants are provided with a list of digits and must fill in the blank spaces with the correct symbols as quickly as possible. The score is determined by the number of correct substitutions made within the allotted time, offering a quantitative measure of the individual's cognitive processing speed and accuracy.

## 2.1 Strength and limitations

One of the primary strengths of the DSST is its simplicity and ease of administration, requiring minimal equipment and training. Its sensitivity to changes in cognitive processing speed and executive functions makes it a valuable addition to cognitive assessment batteries. Additionally, the DSST's ability to provide a quantitative score facilitates the tracking of cognitive changes over time and the evaluation of treatment efficacy.

Despite its advantages, the DSST has limitations. It is influenced by factors such as education and cultural background, which can affect performance and interpretation of results. Moreover, the DSST primarily focuses on processing speed and may not fully capture other cognitive domains such as memory and language. Therefore, it is often used in conjunction with other cognitive tests to provide a more comprehensive assessment of cognitive function.

## 2.2 Applications of DSST

The DSST has been instrumental in aging research, providing insights into the cognitive changes that accompany the aging process [24-26]. By comparing DSST scores across different age groups, researchers have been able to identify patterns of cognitive decline and pinpoint the age at which these declines become more pronounced. Such studies are crucial for developing strategies aimed at preventing or slowing down cognitive aging, enhancing the quality of life for older adults.

Emerging research has utilized the DSST to examine the relationship between lifestyle factors and cognitive function. Studies have explored how physical activity, diet, sleep quality, and social engagement influence DSST performance, shedding light on the potential protective effects of a healthy lifestyle against cognitive decline. For example, DSST scores can predict functional independence and quality of life in older adults. Lower scores are associated with a higher risk of disability, dependency, and poorer health outcomes [27]. Moreover, a study found that obstructive sleep apnea is associated with cognitive impairment in patients with minor ischemic stroke, with the DSST being used to assess cognitive function. This research links sleep disorders with cognitive health [28, 29].

# 3. Montreal Cognitive Assessment (MoCA)

The MoCA is a one-page 30-point test that takes approximately 10-15 minutes to administer. It was introduced by Nasreddine et al. in 2005 [10] to address some limitations of the previously established MMSE, particularly its insensitivity to mild cognitive impairment. The MoCA assesses several cognitive domains including attention and concentration, executive functions, memory, language, conceptual thinking, calculations, and orientation. The primary objective of the MoCA is to detect early signs of cognitive decline in individuals who might present with normal MMSE scores but still experience subtle cognitive deficits that could interfere with their daily functioning.

## 3.1 Strength and limitations

The MoCA is suitable for comprehensive cognitive domains. It evaluates a wide range of cognitive abilities, including executive functions, which are often preserved in early stages of dementia, making it a robust tool for early diagnosis. The test has been validated across a spectrum of neurological conditions, providing reliable assessments for different patient groups [30].

The test requires more time and specific training to administer compared to the MMSE. In addition, MoCA may be subject to cultural and educational bias: scores can be influenced by an individual's education level and cultural background, necessitating adjustments for accurate interpretation [31]. Moreover, in highly educated or high-functioning individuals, the MoCA might not detect very mild impairments [32].

**3.2 Applications of MoCA**

The MoCA is extensively applied in studies investigating cognitive decline and the progression of dementia. Researchers utilize the MoCA to delineate the trajectory of cognitive changes associated with neurodegenerative diseases, such as Alzheimer's disease and related dementias. It is particularly valued for its ability to detect early-stage cognitive impairment that may not be captured by other screening tools [33]. In research focused on neurological disorders like Parkinson's disease, multiple sclerosis, and stroke, the MoCA provides critical insights into cognitive deficits that may emerge. Its sensitivity to a range of cognitive domains makes it a preferred measure for understanding the cognitive sequelae associated with these conditions [34]. In neuroimaging research, the MoCA scores are often correlated with imaging findings to explore structural and functional brain changes in cognitive impairment. This application has contributed to a deeper understanding of the neural mechanisms underlying cognitive decline [35].

# 4. Trail Making Test

The TMT is a two-part pencil and paper test that is easy to administer and is used widely in clinical settings to assess cognitive impairment. The test involves connecting a set of 25 dots as quickly as possible while still maintaining accuracy. The simplicity of the test allows for a quick assessment of a range of cognitive abilities. The part A requires the individual to draw lines sequentially connecting 25 numbered dots (1-2-3, etc.). The task primarily assesses processing speed and visual search abilities, as it requires the individual to identify and connect numbers in order. In the part B, the individual must alternate between numbers and letters (1-A-2-B, etc.). This requires the test-taker to shift back and forth between two sequences, which engages cognitive flexibility and executive control processes [36-38].

## 4.1 Strength and limitations

Integrating the Trail Making Test (TMT) into cognitive assessments provides a nuanced understanding of an individual's cognitive capabilities by measuring processing speed and executive functions. Its comprehensive nature, assessing both straightforward sequencing tasks in Part A and more complex task-switching abilities in Part B, renders the TMT an invaluable tool within a singular evaluative framework. This versatility is particularly beneficial in diagnosing a wide array of neurological disorders, such as Alzheimer's disease and traumatic brain injury, where it has proven diagnostic value. The test's sensitivity to cognitive changes over time enhances its utility in longitudinal studies, allowing for the detailed monitoring of cognitive trajectories. Moreover, its brevity and simplicity of administration make it an attractive option in both clinical and research settings, facilitating widespread usage without the need for extensive resources or time investments.

Despite its strengths, the TMT's interpretative validity can be compromised by external factors like motor speed and physical disabilities, which might affect the outcomes independently of cognitive function. The test's design, which relies on the familiarity with the Roman alphabet and numerical sequencing, introduces potential biases against individuals from varying educational backgrounds and cultures [15, 39]. The pressure of the timed aspect can additionally influence performance, with anxiety potentially detracting from an accurate representation of cognitive capacity. Furthermore, the TMT can exhibit ceiling effects in individuals with high cognitive baseline functioning and floor effects in those with severe cognitive impairments, complicating the assessment of cognitive abilities at these extremes [40-42]. These limitations necessitate a careful approach to interpreting TMT results, especially when considering its application across diverse populations.

## 4.2 Applications of Trail Making Test

Research leveraging the TMT has significantly contributed to our understanding of cognitive aging and the early detection of neurodegenerative diseases, such as Alzheimer's disease. The TMT's sensitivity to executive functions and processing speed makes it a critical tool in distinguishing between normal aging trajectories and the preliminary stages of cognitive decline [11]. This distinction is vital for devising early intervention strategies and for charting the progression of cognitive impairments, offering a clearer picture of the neurodegenerative process.

Moreover, the TMT has been extensively used to assess the efficacy of various cognitive interventions. Studies examining the cognitive benefits of physical exercise, cognitive therapy, and pharmacological treatments have employed the TMT as a measure of improvement in cognitive functions [43]. These investigations provide empirical support for the effectiveness of such interventions in enhancing cognitive health and staving off cognitive decline, highlighting the TMT's relevance in intervention research.

Cross-cultural studies utilizing the TMT have also illuminated the impact of cultural and educational backgrounds on cognitive performance. By comparing TMT results across different cultures, researchers have gained insights into how cultural contexts influence cognitive health and aging patterns [44]. This research underscores the need for culturally adapted cognitive assessment tools and underscores the universality and variability of cognitive aging processes.

# Discussion

The review of the Mini-Mental State Examination (MMSE), Digit Symbol Substitution Test (DSST), Montreal Cognitive Assessment (MoCA), and the Trail Making Test (TMT) underscores the critical nuances each tool brings to cognitive assessment. The MMSE, with its historical precedence and widespread application, provides a valuable starting point for cognitive screening, yet its limitations in sensitivity to mild cognitive impairment (MCI) highlight the necessity for more nuanced tools like the MoCA, which offers a broader evaluation of cognitive domains, thereby facilitating early detection of cognitive decline [10, 45, 46].The DSST's emphasis on processing speed and executive functions complements the MMSE and MoCA, illustrating the importance of assessing specific cognitive abilities in predicting functional outcomes in older adults [9]. Furthermore, the TMT's utility in distinguishing between normal aging and neurodegenerative processes, through its assessment of executive function and processing speed, adds another layer of depth to cognitive evaluations [11].

However, these assessments are not without their limitations. Factors such as cultural and educational background significantly impact performance on these tests, necessitating careful consideration and potential adjustment of normative data to ensure accurate interpretation across diverse populations [31, 47]. Moreover, the inherent challenge in distinguishing between age-related cognitive decline and early stages of neurodegenerative diseases calls for a multifaceted approach to cognitive assessment, combining tools like the MMSE, DSST, MoCA, and TMT with clinical judgment and additional diagnostic procedures [33, 48-50].

Future research should focus on the development and validation of assessment tools that are sensitive to the earliest signs of cognitive impairment, while also being adaptable to the cultural and educational diversity of the aging population. The integration of technological advancements in cognitive testing presents a promising avenue for enhancing the accuracy and accessibility of cognitive assessments, potentially allowing for more widespread screening and early detection of cognitive decline.

The ongoing refinement of cognitive assessment tools, informed by research and clinical practice, will be crucial in addressing the growing challenge of cognitive impairment and dementia in an aging global population.

## Acknowledgments

We thank Dr. Chao Qin of University of Southern California for critically reviewing the manuscript and for offering significant insights.

## Declaration of interests

The authors declare no competing interests.